\documentstyle[epsf]{mn}
\title{\bf Dark galaxies, spin bias and gravitational lenses} 

\author[R. Jimenez et al.]{R. Jimenez$^1$, A.F. Heavens$^2$, 
M.R.S. Hawkins$^1$, P. Padoan$^3$ \\
$^1$Royal Observatory, Blackford Hill, Edinburgh EH9 3HJ, U.K. \\
$^2$Institute for Astronomy, University of Edinburgh, 
Blackford Hill, Edinburgh EH9 3HJ, U.K. \\
$^3$Theoretical Astrophysics Center, Juliane Maries Vej 30, 
DK-2100 Copenhagen 0, Denmark \\}
\newcommand{\ls}{\raisebox{-.8ex}{$\buildrel{\textstyle<}\over\sim$}}
\newcommand{\gs}{\raisebox{-.8ex}{$\buildrel{\textstyle>}\over\sim$}}
\begin{document}
\maketitle

\begin{abstract}
Gravitational lensing studies suggest that the Universe may 
contain a population of dark galaxies;  we investigate this 
intriguing possibility and propose a mechanism to explain their nature. 
In this mechanism a dark galaxy is formed with a low density disk in a  
dark halo of high spin parameter;  such galaxies can have surface densities
below the critical Toomre value for instabilities to develop, and 
following Kennicutt's work we expect these galaxies to have low 
star formation rates.  The only stellar component of the galaxies is a halo
system, formed during the collapse of the proto-galactic cloud. 
We compute synthetic stellar population 
models and show that, at a redshift $z=0.5$,  such galaxies have 
apparent magnitudes $B \simeq 28, R \simeq 26$ and $I \simeq 25$, 
and could be unveiled by deep searches with the Hubble Space Telescope.  
Dark galaxies have an initial short blue phase and then become 
essentially invisible, therefore they may account for the blue population 
of galaxies at high redshift.
We find a strong mass-dependence in the fraction of dark galaxies,
and predict that spiral galaxies will not be found in halos with masses 
less than about $10^9\,M_\odot$, if $\Omega=1$.  Above about $10^{12}M_\odot$, 
all halos can produce luminous disks.  The mass-dependence of the 
galaxy-formation efficiency introduces the possibility of `spin bias' --
luminous galaxies being associated preferentially with strongly-clustered
high-mass halos.  A further prediction is that the slope of
the faint-end luminosity function for galaxies will be flatter than the 
associated halo mass function.
 
\end{abstract}

\begin{keywords}
cosmology: theory --- cosmology: dark matter --- galaxies: formation -
-- galaxies: evolution
\end{keywords}

\section{Introduction}

The first gravitational lens was discovered by Walsh, 
Carswell \& Weymann (1979),  and
comprised a pair of quasar images split by a clearly visible massive galaxy.
Since then, many manifestations of gravitational lensing have been observed,
including systems of two or four images, or a ring.  Although in some cases
a lensing galaxy can be seen, over the last few years extensive evidence has
been accumulating that multiple and complex quasar images are being
gravitationally lensed by dark objects of galaxy mass, and the phrase `dark
galaxy' has frequently been used to characterise such bodies (Weedman et al., 
1982, Djorgovski \& Spinrad 1984, Meylan \& Djorgovski 1989,
Hewett et al. 189, Wisotski et al. 1993, Hawkins et al. 1997). 

Each configuration of quasar images presents its own particular problems in 
interpretation, and it is probably fair to say that any particular system 
can be dismissed on the basis of an improbable conjunction of phenomena.  
In the case of double quasars, in spite of arguments based on similarity of 
spectra, colour and redshift, there is still some residual doubt that they 
may be chance coincidences.  In these systems, under the lensing hypothesis 
it is usually possible to put tight limits on the mass-to-light ratio
of the lensing galaxy, typically several hundred $M_{\odot}/L_{\odot}$.
For quadruple systems there is usually no doubt that they are
indeed lensed systems, but the close proximity of the four images can make
it hard to put useful limits on the magnitude of the lensing galaxy.
Potentially the most conclusive case for dark galaxies comes from radio rings.
Here a compact radio source is lensed to produce a ring.  There are several
examples already known where no lensing galaxy is clearly detectable, but
optical emission from the radio source coupled with the small diameter of
the rings tends to complicate the picture.  Although in individual cases
it is often possible to explain away the necessity for a dark galaxy, if
one looks at the population of lensed quasars as a whole, the case for
unseen lenses is hard to avoid.  For example, a sample of double quasars has
recently been analysed by Hawkins (1997), and a strong statistical
case is presented that dark objects of galactic mass must be present in most
systems.

Rather surprisingly, in spite of the conclusion in a number of papers that
lensing is probably being caused by dark galaxies or dark matter halos, little
attempt has been made in the literature to explain the nature of these
invisible gravitational lenses.  Indeed, regardless of the lensing evidence,
the question of whether dark galaxies can exist is an intriguing one
(Dekel \& Silk 1983, Silk 1986).

In this letter we propose a mechanism to explain the nature of dark galaxies.
While it seems difficult to prevent massive galaxies from producing stars on 
initial collapse (Tegmark et al. 1997; Padoan, Jimenez \& Jones 1997), we 
suggest that some galaxies only form halo stellar components, while
their disks remains gaseous and quiescent if the surface density is low 
enough and they do not undergo (further) mergers.  
These galaxies are extreme cases of Low Surface Brightness (LSB) 
galaxies, and we propose to name them Low Density Galaxies (LDG).

We argue that the angular momentum of the galactic halo may play 
an important role in determining whether galaxies become 
LDGs and therefore dark.  Scaling computations show that the possibilities 
for dark galaxies are higher for lower-mass halos, and we predict a minimum
mass of halos with bright disks of around  $10^9 M_\odot$ in an Einstein-de 
Sitter Universe.  Dark lenses of high mass $\sim 10^{12} M_\odot$
would be the largest predicted by this theory.  Nevertheless,
there are important general implications of this work.  Apart from the
minimum spiral mass already mentioned, the predicted
increasing fraction of dark galaxies towards lower masses modifies the
luminosity function so it has a flatter faint slope
than the associated halo mass function;  the initial burst of star formation
may provide a population of `disappearing dwarfs' proposed for interpreting the
faint galaxy counts (cf Babul \& Rees 1992, Metcalfe et al. 1996). 
Furthermore, for intermediate mass galaxies, the 
correlation between bright disks and halo mass provides a mechanism for 
biasing the luminous galaxy population 
towards high-density regions as it is known that more massive halos
are expected to inhabit preferentially 
high-density regions (Cole \& Kaiser 1989, Mo \& White 
1996)\footnote{With the exception of this footnote, 
in this paper bias always has the cosmological meaning of the different 
clustering properties of mass and galaxies, not the observational bias in 
favour of low-spin systems implied by this analysis}.   

\section{Low density galaxies and the spin parameter}

It is known that some galaxies transform virtually all their gas into 
stars and others do not. The case of LSB galaxies is a good 
example where star formation in the disk has been inhibited, due to the low 
surface density of the disk (van der Hulst et al. 1993). These galaxies 
form most of their stars in a first burst and then evolve into a 
rather quiescent phase (Padoan, Jimenez \& Antonuccio-Delogu 1997). 
On the other hand, High Surface Brightness (HSB) galaxies continually form 
stars, because of their higher surface density $\Sigma$. 

The surface density in a galaxy disk and star formation 
activity can be related using the Kennicutt (1989) formalism,
based on the Toomre (1964) stability criterion.  If 
the Toomre parameter $Q=\Sigma_c/\Sigma$, where $\Sigma_c$ is a critical 
surface density, exceeds about unity, very little star formation takes place.

We assume that disks settle until they are rotationally-supported 
(cf Fall \& Efstathiou 1980, White \& Rees 1978), so the final surface
density is related to the initial spin parameter of the halo.  If the 
spin parameter of the halo is large enough, $Q>1$ and it is assumed that 
no star formation takes place in the disk.

The spin parameter of dark matter halos has been studied analytically and
numerically by a number of authors (e.g. Heavens \& Peacock 1988, 
Barnes \& Efstathiou 1987,  Warren et al. 1992, Catelan \& Theuns 1996), and
is expected to anti-correlate only weakly with peak height (and hence 
environment), and the range of spin parameters is broad 
$0.01 \ls \lambda \ls 0.1$.

In this section, we calculate the settling radius and final surface density 
of the gas, and compare with the Toomre (1964) criterion for instability of the
disk.  It is difficult to make the
argument rigorously quantitative, because the effects of sub-clumping and 
interaction between the baryonic component and dark matter may not be trivial.
However, the general picture may be illustrated by a model in which the
gas settles in a (given) dark matter potential, until it is 
rotationally-supported.   For small baryon fractions, the dynamics are
still dominated by the halo (see below and Dalcanton et al. 1997),
especially for the extended disk systems which are of most interest in
this paper.

We assume that the gas component has the same 
specific angular momentum as the dark matter initially, both arising 
because of tidal torques (see Navarro \& Steinmetz for an alternative view).  
We take the universal dark matter 
profile found in simulations by Navarro, Frenk \& White (1996):
\begin{equation}
{\rho(r)\over \rho_{\rm crit}} = {\delta_c r_s^3 \over r(r+r_s)^2}
\end{equation}
where $r_s \sim 5$ kpc is the `core' radius of the halo, containing
typically around 10\% of the mass.  $\delta_c \sim 5 \times 10^3\Omega 
(1+z_f)^3$ is
the characteristic density in units of the background critical density 
(White 1997). 
The bounding radius of the virialised halo, $r_{200}$  is defined as the 
radius within which the average density is 200 times the critical density,
and is typically about $10r_s$.

The spin parameter is 
$\lambda = J|E|^{1/2}/GM^{5/2}$, where $J$ is the angular momentum of the 
halo and $E$ its energy.  We assume that the dark matter is the dominant 
contributor to the potential,  and compute the settling radius of baryonic
matter of initial projected radius $r_i$.  Equilibrium of the 
gas in the dark matter halo is achieved at a radius $r$ given by the
solution to the equation $r M(<r)= a^2\lambda^2 r_i M(<r_i)$, 
where $a \sim 2.5$,  defined such that the square of the halo
rotation speed at $r_i$ is $a^2 \lambda^2 G M(<r_i)/r_i$, and the
value of $a$ corresponds to $\lambda \sim 0.4$ for rotational support 
(White 1997).  
The surface density of the baryonic disk is given by $r \Sigma(r) = \Sigma(r_i)
r_i {\rm d}r_i/{\rm d}r$, where the initial baryon surface density follows
the dark matter profile, but scaled by the ratio $\Omega_B/\Omega$.  We 
take the baryon contribution to the density parameter $\Omega$ to be 
$\Omega_B=0.035$, consistent with primordial
nucleosynthesis if $H_0=65$ km\,s$^{-1}$ Mpc$^{-1}$ (Walker
et al. 1991), so the self-gravity of the disk makes a small correction 
to the settling radius.  The final surface density profile is shown for
two halo masses in Fig.~1.   Note that 8 percent of the halo mass resides 
within a radius of $0.2 r_s$,  so the dark matter still dominates there.

The mean value of the spin parameter, $\simeq 0.05$ and its dispersion 
$\sim 0.03$ are only weakly dependent on the power spectrum 
(Heavens \& Peacock 1988, Barnes \& Efstathiou 1987),  
so conclusions we draw from Fig. 1 are not model-sensitive.  
Also, the trends shown in Fig.~1 -- larger gas distributions
and lower densities for larger halo spin parameters, are clearly general
features and not specific to the assumptions made here.  

We turn now to the quantitative question of whether the resulting surface
density is above or below the critical value apparently required for
star formation.  The critical surface density
is $\Sigma_c =  \alpha \kappa c_{disp}/(3.36 G)$ (Kennicutt 1989), 
where $\alpha \simeq 1$, $c_{disp}$ is the 
velocity dispersion of the disk, and the epicyclic frequency,
written in terms of the rotation speed, is $\kappa = 1.41 (V/r)
\sqrt{1+{\rm d}{\ln V}/{\rm d}\ln r}$.
The Toomre stability parameter $Q \equiv \Sigma_c/\Sigma$ may then be written
in terms of $s\equiv r/r_s$, exploiting the scalings in the problem
(ignoring a weak mass-dependence of $\delta_c$ (White 1997)):
\begin{eqnarray}
Q(s) = {\alpha\over 3.36} \sqrt{{\pi\over G \delta_c\rho_{crit}}} 
{c_{disp}\Omega\over 
r_s\Omega_B} \times \nonumber \\ 
{f\over s^{3/2}I(s/f)} {df\over d\lambda} 
\sqrt{M(s)+{s^2\over(1+s)^2}}.
\end{eqnarray}
where $f(s,\lambda)=r/r_i$ is the solution to $sM(s)=a^2\lambda^2 s_i M(s_i)$;
the enclosed mass is related to $M(s)\equiv\ln(1+s)-s/(1+s)$, and the 
initial surface density is related
to $|s^2-1|^{3/2}I(s) = \sqrt{s^2-1}-\cos^{-1}(1/s) (s>1)$ and 
$=-\sqrt{1-s^2}+\ln\left[s(1-\sqrt{1-s^2})^{-1}\right] (s<1)$.
  
In Fig. 1 and 2 we show the $Q$ parameter for two
halos of different mass, $M=10^8 M_\odot$ and
$M=10^{12} M_\odot$.   The
halos have formation redshift 1, but the 
curves are not very sensitive to these parameters.  Following
Bahcall and Casertano (1984), we take $c_{disp}=6$ and 15 km\,s$^{-1}$ 
respectively, 
since we have no particular reason to believe that the velocity dispersion 
should depend sensitively on the star formation rate.  Eleven curves are
shown, corresponding to halos with $\lambda$ in the expected range 
between $0.01$ and $0.1$.  As expected, the higher spin halos are 
more stable, and the low-mass halo is stable for more-or-less all 
reasonable spin parameters.  The high-mass halos, on the other hand,
all produce stars somewhere in the disk.  We do not consider here 
bar instabilities (e.g. Christodoulou et al. 1995), but note that
the fastest-rotating ($\lambda \gs \Omega_B/\Omega$) disks should
be stable (Mo et al. 1997). 

\begin{figure}
\begin{center}
\setlength{\unitlength}{1mm}
\begin{picture}(90,70)
\includegraphics{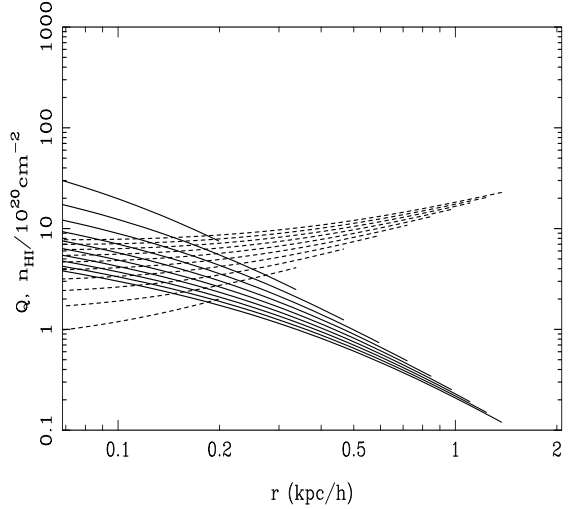}
\end{picture}
\end{center}
\caption{(Solid line) 
The gas surface density (in hydrogen atoms per square cm, divided by 
$10^{20}$),  for spin parameters from $0.01$ (top) to $0.1$ (bottom), in
steps of $0.01$.  The halo mass is $10^8 M_\odot$, the formation
redshift $z_f=1$, and $\Omega=1$.  The dashed line shows the Toomre stability 
parameter $Q$.  A disk with $Q$ in excess of about unity is theoretically 
stable, and 
observationally associated with very little star formation.}
\end{figure}

\begin{figure}
\begin{center}
\setlength{\unitlength}{1mm}
\begin{picture}(90,70)
\includegraphics{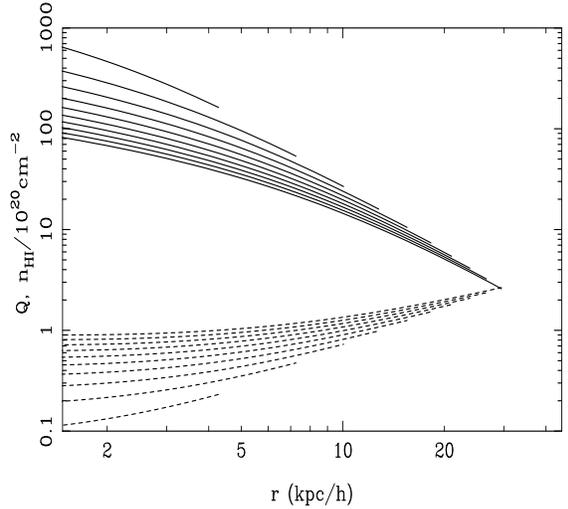}
\end{picture}
\end{center}
\caption{As Fig. 1, but $M=10^{12} M_\odot$.}
\end{figure}

In Fig. 3 and 4, we summarize the results as follows.  For each halo mass and 
formation redshift, we calculate the maximum spin parameter which will 
produce stars somewhere in the disk ($Q<1$).   Fig. 3 shows results
for $c_{disp}=10 km s^{-1}$; for other values scale the mass by the cube
of $c_{disp}/10$.   We see that the dependence
on formation redshift is weak, and that higher mass halos have a progressively
larger critical $\lambda$.  In Fig. 4, we show the fraction of halos giving
rise to luminous disks in the model, assuming a gaussian distribution
for $\ln(\lambda)$ with mean $\ln(0.05)$ and dispersion 0.5 (Mo et al. 1997). 
The solid curve is for $c_{disp}=
10$ km s$^{-1}$, the dashed curve allows a smooth mass-dependence of $c_{disp}$
from 5 to 15 km s$^{-1}$ as the mass changes from $10^8$ to $10^{12} M_\odot$.
The fraction of luminous halos increases from 5\% at $2 \times 10^9 M_\odot$ 
to 95\% at just over $10^{12} M_\odot$.

\begin{figure}
\begin{center}
\setlength{\unitlength}{1mm}
\begin{picture}(90,70)
\includegraphics{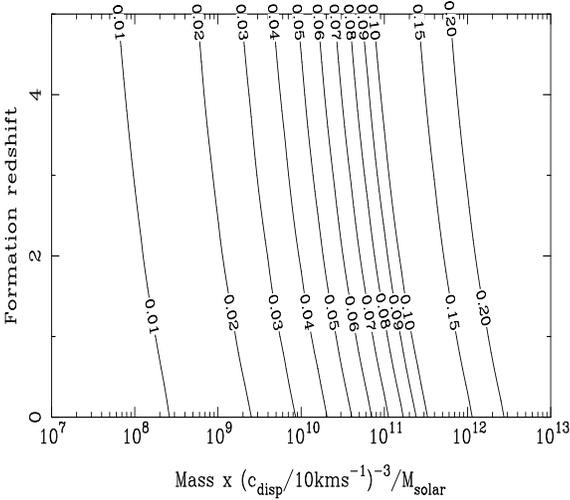}
\end{picture}
\end{center}
\caption{The maximum halo spin parameter which will produce stars 
somewhere in the disk, according to the Toomre stability criterion,
for halos of different total mass and formation redshift.  The 
expected range of spin parameters for halos is $0.01-0.1$.  The
assumed disk velocity dispersion here is $10 km s^{-1}$.
Scaling solutions imply that the masses here scale as $c_{disp}^3$.}
\end{figure}

\begin{figure}
\begin{center}
\setlength{\unitlength}{1mm}
\begin{picture}(90,70)
\includegraphics{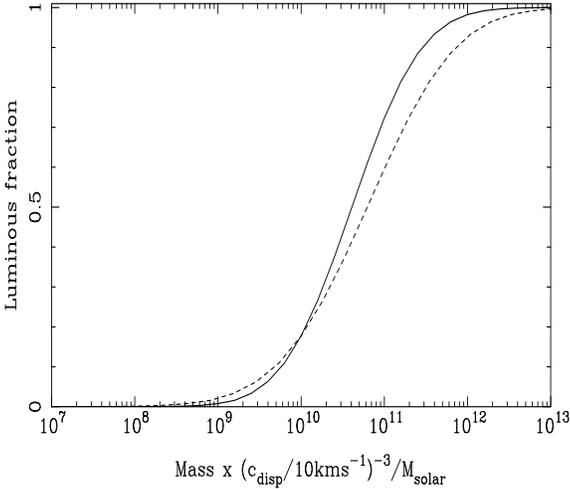}
\end{picture}
\end{center}
\caption{The fraction of halos producing luminous disks as a function of
total halo mass.  The solid curve assumes a disk velocity dispersion of
10 km s$^{-1}$, the dashed curve assumes the dispersion increases 
linearly with $\log(M)$ as indicated in the text.}
\end{figure}

Dark gravitational lenses of high mass $\sim 10^{12} M_\odot$, apparently 
required to explain the missing lens double quasar population (Hawkins 1997),
are possible as the most massive examples of halos with dark 
rotationally-supported disks.     These halos always have the
possibility to cause splittings of images;  the singular density profile
assures this,  and splittings of up to about 4 arcsec are possible from
$10^{12} M_\odot$ halos at $z \sim 0.5$.

In addition to the dark lensing possibility,
there are many other interesting consequences of this model:  
the mass-dependence of the 
bright fraction will alter the luminosity function -- it will no longer
reflect the underlying halo mass function, which may be useful since most 
currently popular galaxy formation models over-predict the low-mass population;
perhaps most interestingly,  there should be no spiral galaxies with total 
masses below about $10^9 M_\odot$ (cf. compilation by Roberts \& 
Heynes 1994);    low-mass or high-spin halos should lead to 
short-lived bright galaxies,  affecting the interpretation of the faint 
blue number counts;  luminous galaxies should 
be found in high-density environments.  The last two points are explored 
in the next two sections.

It should be noted that the possibilities for dark galaxies are much lower
if non-baryonic dark matter contributes much less than the critical 
density.    In this case, the minimum spiral mass drops
substantially, to around $10^7 M_\odot$ if $\Omega_0=0.2$, but the 
disk self-gravity cannot be ignored in this case.

\section{Star formation in low density galaxies}

LDGs will not be entirely dark, as it seems impossible to prevent
some fragmentation of the initial halo (Tegmark et al. 1997; Padoan, 
Jimenez \& Jones 1997).   Without ongoing star formation in the disk, the
galaxy will rapidly fade,  producing a short-lived bright population of
`disappearing dwarfs'.   The details of the process are not crucial for
the arguments in this paper, but for illustration we calculate the
brightness of a halo of mass $10^{10} M_\odot$, using the star formation
model of Padoan et al. 1997, with formation redshifts of 0.5 and 1 (Fig. 5).
We also show the colour evolution in Fig. 6.    The assumptions here
are 2\% star formation efficiency (a prediction of the model, but the
results may be scaled if desired), and an
initial metallicity of Z=0.002, Y=0.24.  The graphs use the latest 
version of our synthetic stellar population code 
(Jimenez et al. 1997).    
These results evidently have consequences for the number counts of galaxies 
since the initial bright phase of LDGs may account for the population of 
blue galaxies found at 
high redshifts (cf Metcalfe et al 1997 and references therein).
In the most extreme cases (Tyson et al. 1986) the observed limiting 
magnitudes for dark lenses are: 
$I > 24.5$, $R > 26$ and $B > 25$. And for the surface brightness $R > 27.9 
\rm\, mag\, arcsec^{-2}$. Therefore, even at a redshift of 0.5, a LDG has a 
magnitude 
below the limits found by (Tyson et al. 1986), already 2-3 Gyr after 
the formation of its 
halo, and its mass-to-light ratio is about $1000M_{\odot}/L_{\odot}$, as 
required.

\section{Spin Bias}

The possible existence of dark galaxies has implications for
bias.  A bias in the luminous galaxy population has been invoked as a 
way to reconcile an Einstein-de Sitter universe with observations of
low mass-to-light ratios in galaxy clusters, provided that luminous 
galaxies are found preferentially in clusters.  On larger scales, a 
bias $b$ has important implications for determining the density parameter from
peculiar velocity studies (e.g. Willick et al. 1997) or redshift-space
distortions (e.g. Heavens \& Taylor 1995), since linear studies determine
$ \Omega$ only in the combination $\beta \equiv \Omega^{0.6}/b$.  
Results from these studies are inconclusive at present, with values of
$\beta$ ranging between about 0.5 and 1 (see e.g. Strauss \& Willick 1995).

Spin bias will act in a similar way to high peak biasing, or natural biasing
(Kaiser 1984, Davis et al. 1985, Bardeen et al. 1986,  
Cole \& Kaiser 1989, Mo \& White 1996), where lower-mass halos are 
anti-correlated with respect to higher-mass halos.  The link with observation
then requires a formula for determining which halos result in luminous 
galaxies.  Spin bias provides such a link,  and the efficiency of luminous
disk formation will preferentially select the higher-mass halos, which are 
biased with respect to the dark matter (cf. Kauffmann et al. 1997).

\section{SUMMARY}

In this letter we have proposed a new scenario to explain the nature of dark 
lenses based on Low Density Galaxies, that we define as galaxies
with a stellar halo component, and a rotationally supported gaseous 
disk unable to form stars because of its low surface density, which 
is mainly a consequence of the large spin parameter of the galactic halo. 
%These galaxies 
%are extreme cases of Low Surface Brightness (LSB) galaxies, as described in 
%Padoan, Jimenez \& Antonuccio-Delogu (1997).

\begin{figure}
\centering
\leavevmode
\epsfxsize=1.0
\columnwidth
\epsfbox{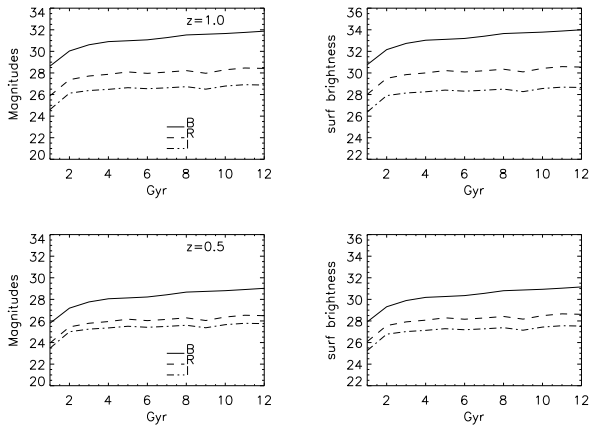}
\caption[]
{The panel shows the predicted magnitudes and surface brightness 
at different redshifts ($z=0.5$ and $z=1.0$) for the model of a galaxy with 
low efficiency (2\%) in star formation (LDG). The LDG is below the observed 
limits (see text).}
\end{figure}

The model with which we have chosen to demonstrate the ideas
assumes a particular form of dark matter halo, and a specific star formation
model to obtain the brightness of the galaxy, but these ingredients are not
central to the argument.   The general feature of high-spin systems with little
initial star formation producing rotationally-supported spirals with large
radii is robust.  If one accepts the Kennicutt criterion for star 
formation,  the possibility of dark galaxies then emerges quite naturally.  
Support for the relevance of the Kennicutt
criterion comes from LSB galaxies, which show very little current star
formation activity.
Our analytic calculations can, of course, be regarded only as an approximation
in hierarchical models of galaxy formation, and future gas-dynamic modelling
will provide a more robust test of the hypothesis.  

A firm prediction of our model is that LDGs should be seen in 
deep HI surveys in voids. 
The column density of LDGs can be high ($\sim 10^{21}$ cm$^{-2}$ but 
mass-dependent), but there is the possibility of ionization of the hydrogen,
which would prevent their appearance in HI surveys.   In any case, due 
to the way HI surveys are performed, there could be many
undetected massive HI objects within a distance of $250h^{-1}$ Mpc
(Briggs 1990), but LDGs would be revealed by next generation HI 
surveys (e.g. Parkes)\footnote{After this paper was completed, our attention 
was drawn to a paper by Meurer et al. (1996), in which they report the
discovery of a very dark galaxy, with $M/L=79$, a large HI disk,  
and star formation 
confined to the inner 0.8 kpc.  This fits well with our predictions, being a 
transitional case of a galaxy which is unstable only in the central parts.
The authors suggest that the star formation is controlled principally by the
column density $n_{HI}>10^{21}$ cm$^{-2}$, rather than $Q$, but the qualitative
effect is the same, as seen in Fig. 1 and 2.}

The main conclusions of this work are:

\begin{itemize}

\item The spin parameter of the halo may play a critical role in 
determining whether galaxies quickly become dark.

\item Spiral galaxies should have masses in excess of about $10^9 M_\odot$,
and there is an upper limit for dark galaxies of around $10^{12} M_\odot$,
if $\Omega_0=1$.

\item LDGs convert about 2\% of their baryonic mass into stars in the first 
couple of Gyr of their formation process. They consist of a stellar halo 
and a gaseous, quiescent disk.
LDGs have apparent magnitudes $B \simeq 28$ mag at redshift  z=0.5 and  
$B \simeq 32$ mag at z=1.0; they are therefore invisible.
The colours of LDGs are very similar to the colours of LSB galaxies. 

\item A large fraction of the total number of galactic halos
might contain an invisible LDG, with a higher probability in low mass 
halos, thus providing a mechanism, `spin bias', for biasing the luminous
galaxy distribution.

\item The faint-end luminosity function of spiral galaxies should be 
flatter than the corresponding halo mass function, and there should be 
a population of `disappearing dwarfs'.

\end{itemize}

\begin{figure}
\centering
\leavevmode
\epsfxsize=1.0
\columnwidth
\epsfbox{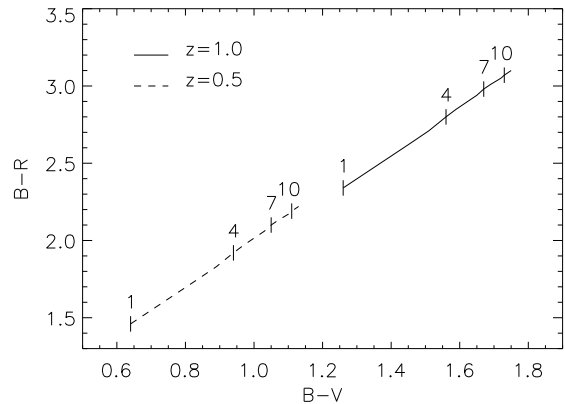}
\caption[]
{The predicted colours (B-V vs B-R) for LDGs at redshifts $z=0.5$ 
(dashed line) and $z=1.0$ (continuous line). The numbers on the tick marks 
correspond to the age of the theoretical model in Gyr. 
 The colours of LDGs are similar to those found for LSBs.}
\end{figure}

\section*{acknowledgments}

The authors thank the referee for comments which led to a substantial
improvement in this paper, and Chris Flynn for drawing our attention
to the near-dark galaxy NGC2915.

\end{document}